\documentclass[twocolumn, preprintnumbers, prl]{revtex4-1}
\usepackage{amsmath}
\usepackage{amsfonts}
\usepackage{amssymb}
\usepackage{amsthm}
\usepackage{bm}
\usepackage{graphicx}
\usepackage{url}
\usepackage{hyperref}

\usepackage{lmodern}
\usepackage{subfigure}
\usepackage[section]{placeins}
\usepackage{xcolor}
\usepackage{float}
\usepackage{multirow}
\usepackage[utf8]{inputenc}
\usepackage[T1]{fontenc}
\usepackage{hyperref}

\newcommand{\bi}{\begin{itemize}}
\newcommand{\ei}{\end{itemize}}
\newcommand{\non}{\nonumber}
\newcommand{\bea}{\begin{eqnarray}}
\newcommand{\eea}{\end{eqnarray}}
\newcommand{\be}{\begin{equation}}
\newcommand{\ee}{\end{equation}}
\newcommand{\ben}{\begin{eqnarray*}}
\newcommand{\een}{\end{eqnarray*}}
\newcommand{\bem}{\begin{pmatrix}}
\newcommand{\eem}{\end{pmatrix}}
\newcommand{\bl}{\begin{align}}
\newcommand{\el}{\end{align}}
\newcommand{\beg}{\begin{gather}}
\newcommand{\eeg}{\end{gather}}

\newcommand{\mi}{\mathrm{i}}
 
\newcommand{\cD}{\mathcal{D}}

\newcommand{\cF}{\mathcal{F}}

\newcommand{\cH}{\mathcal{H}}

\newcommand{\cN}{\mathcal{N}}

\newcommand{\bM}{\ensuremath{\mathbb{M}}}

\newcommand{\bR}{\ensuremath{\mathbb{R}}}

\newcommand{\bZ}{\ensuremath{\mathbb{Z}}}

\newcommand{\apm}{\alpha'}

\newcommand{\IH}{\mathbb{H}}

\newcommand{\e}{\epsilon}

\renewcommand{\r}{\rho}                                     
                                   
\renewcommand{\t}{\tau}

\newcommand{\vt}{\vartheta}

\newcommand{\TrH[1]}{ {\raise -.5em
                      \hbox{$\buildrel {\textstyle  {\rm Tr } }\over
{\scriptscriptstyle \cH _ {#1}}$}~}}
\newcommand{\res[1]}{{\raise -.5em 
\hbox{$\buildrel{\textstyle{\rm Res}}\over {\scriptscriptstyle {#1}}$}}}
\newcommand{\tends[1]}{{\raise -.5em 
\hbox{$\buildrel{\longrightarrow}\over {\scriptscriptstyle {#1}}$}}}

\newcommand{\half}{\frac{1}{2}}

\newcommand{\Tr}{\mbox{Tr}}

\renewcommand{\Re}{\mbox{Re}}

\def\dbend{\lower3.5pt\hbox{\manual\char127}}

\def\IL{\relax{\rm I\kern-.18em L}}
\def\IH{\relax{\rm I\kern-.18em H}}
\def\rlx{\relax\leavevmode}

\def\ZZ{\rlx\leavevmode\ifmmode\mathchoice{\hbox{\cmss Z\kern-.4em Z}}
 {\hbox{\cmss Z\kern-.4em Z}}{\lower.9pt\hbox{\cmsss Z\kern-.36em Z}}
 {\lower1.2pt\hbox{\cmsss Z\kern-.36em Z}}\else{\cmss Z\kern-.4em
 Z}\fi}

\begin{document}
\title{Finite Entanglement Entropy in String Theory}
\author{Atish  Dabholkar} 
\author{Upamanyu Moitra}
\affiliation{Abdus Salam International Centre for Theoretical Physics\\
Strada Costiera 11, Trieste 34151 Italy}

\begin{abstract}
We analyze the one-loop quantum entanglement entropy in ten-dimensional Type-II string theory using the orbifold method by analytically continuing in $N$ the genus-one partition function for string orbifolds on $\mathbb{R}^2/\mathbb{Z}_N$ conical spaces known for all odd integers $N > 1$. We show that the tachyonic contributions to the orbifold partition function  can be appropriately summed and analytically continued to an expression that is finite in the physical region $0 <  N \leq 1$ resulting in a finite and calculable answer for the entanglement entropy. We discuss the implications of the finiteness of the entanglement entropy for the information paradox, quantum gravity, and holography.
\end{abstract}

\maketitle

\subsection{Introduction}

Entanglement entropy is a quantity of fundamental importance in quantum mechanics and quantum field theory, and even more so in quantum gravity. The naively defined von Neumann entropy measuring the entanglement between the inside and the outside of  a black hole is divergent in quantum field theory and proportional to the horizon area in units of the short-distance cutoff. This divergence across any sharp boundary is  a consequence of the fact that field values on the two sides of the boundary have strong short-distance correlations in a local quantum field theory. If this divergence is not cured in quantum gravity, then it would imply that the black hole has infinite number of qubits and can store arbitrary amount of information. Unitary evolution would then be impossible unless the black hole is interpreted as a remnant with all the attendant problems of this interpretation. Finiteness of entanglement entropy is thus at the heart of the information paradox in black hole physics. 

Given the ultraviolet finiteness of string perturbation theory, it behooves us to ask if one can define a suitable notion of entanglement entropy in string theory and examine its finiteness order by order. A direct definition of such a quantity has proven elusive partly because it is not clear how to define in string theory the relevant density matrix and appropriate notions corresponding to its von Neumann or R\'enyi entropy. One expects that it would be difficult to introduce sharp boundaries in string theory given the soft behavior of strings at short distances.   One can instead attempt an indirect definition by a generalization of  R\'enyi entropy adapted to string theory using $\mathbb{Z}_N$ orbifolds \cite{Dabholkar:1994ai}. The simplest example constructed in \cite{Dabholkar:1994ai} is Type-II
 string 
 on $\bM_8 \times \mathbb{R}^2/\mathbb{Z}_N$ where $\bM_{8}$ is $7+1$ dimensional Minkowski spacetime, $\mathbb{R}^2$ is two-dimensional Euclidean plane, and the orbifold action is generated by a rotation in the plane through an angle $4\pi/N$ for $N$ odd. 

It is convenient  to write  entanglement  entropy as
\begin{equation}
	S = S^{(0)}	+ S_q
\end{equation}
where $S^{(0)}$ is the classical contribution  and $S_q$ is the quantum contribution from higher-genus  Riemann surfaces \cite{Dabholkar:2001if,  Dabholkar:2022mxo}. 
The classical spacetime partition function   $\hat Z^{(0)}(N)$ of the $\bZ_{N}$ orbifold  theory  is nontrivial and analytic in $N$ after including a boundary contribution, and  $S^{(0)}$ is simply given by the Bekenstein-Hawking entropy  \cite{Dabholkar:2001if} but with tree-level, unrenormalized Newton's constant $G_0$.
The quantum spacetime partition function $\hat Z_q(N)$ is related to the worldsheet partition function   $Z_q(N)$  by
\be\label{spaceworld}
\log (\hat Z_q(N))  = Z_q(N) \,  
\ee
with a genus expansion
\begin{equation}\label{Zq}
	Z_q(N) = \sum_{g=1}^\infty Z^{(g)}(N) \, .
\end{equation}
The expression for $\hat Z_q(N)$  is expressed as a sum over orbifold sectors and is not obviously analytic. If  an analytic continuation exists, then the quantum entanglement entropy would be given \cite{Dabholkar:1994ai, Dabholkar:2001if,  Dabholkar:2022mxo} by a perturbative expansion
\begin{equation}\label{S-total}
	{S}  = \frac{\partial}{\partial N} \left( N \log (\hat Z_q(N) \right) \bigg|_{N=1} = \sum_{g=0}^\infty S^{(g)} \, = \frac{A}{4G_0} + S_q \, .
\end{equation} 
The orbifold method thus applies uniformly for computing both the classical and quantum contributions. 

It is significant that the orbifold method  in string theory automatically supplies a classical term in \eqref{S-total} related to the Bekenstein-Hawking formula, and that all higher order terms are proportional to area. It thus appears to offer a natural gravitational generalization of von Neumann  entropy with a systematic expansion that is intrinsically holographic. The computation is formally analogous to the replica method in field theory \cite{Mezard:1986xw} and \textit{a priori} has nothing to do with black hole physics. But, unlike in field theory, it appears that in a consistent theory of gravity  the classical term is inevitable in a discussion of quantum entanglement. In fact, it is  of crucial importance.  Heuristic arguments indicate that field-theoretic divergences in the quantum entanglement entropy $S_q$ can be absorbed into renormalization of Newton's constant \cite{Susskind:1994sm, Kabat:1995eq} suggesting that the total entropy must be finite. In semiclassical gravity, a corresponding fact in the more abstract formulation is that inclusion of gravity turns the algebra of observables from Type-III to Type-II \cite{Witten:2021unn, Chandrasekaran:2022cip}.  As discussed in \cite{Dabholkar:2022mxo}, one expects a stronger statement that the `algebra of observables' should be akin to Type-I corresponding to finite entanglement entropy.

Since the quantum effective action includes not only local terms but also nonlocal terms arising from loops of massless fields, not all  contributions can be attributed to the renormalization of Newton's constant in the Wilsonian action. Therefore,  the entanglement entropy should be finite but nonzero order by order in string perturbation theory. Holographic considerations discussed in \cite{Dabholkar:2022mxo} also confirm this expectation.  

It is possible that the entropy defined by \eqref{S-total}  is a more fundamental notion than black hole entropy if one can make sense of \eqref{S-total}.  At the quantum level, the definition of black hole entropy has potential ambiguities from the choice of the ensemble and from the contributions of the thermal bath surrounding the black hole. For  supersymmetric black holes at zero temperature it is possible to define and in some cases compute exactly \cite{Dabholkar:2014ema} the quantum entropy, but these ambiguities need to be resolved. Entropy in \eqref{S-total} is free of these ambiguities. It is more general and geometric since it focuses on the bifurcate Rindler horizon which in the Euclidean continuation corresponds to the tip of the cone. It thus depends only on dividing space or the Hilbert space into two parts and not on a specific spacetime geometry or the asymptotics. Moreover, \eqref{S-total} presumably gives the fine-grained entropy much like the von Neumann entropy and not the coarse-grained thermodynamic entropy as for a black hole. 

In this note we develop further the program initiated in \cite{Dabholkar:1994ai}  to compute quantum entanglement entropy using the orbifold partition functions as the starting data, and in particular to examine the finiteness of entanglement entropy. This idea immediately encounters a possible obstacle. The spectrum of the $\mathbb{Z}_N$ orbifold contains several tachyons for all $N>1$. As a result, the worldsheet partition function $Z_q(N)$ suffers from severe infrared divergences \cite{Dabholkar:1994ai}. It would thus appear that we have simply traded the ultraviolet divergence for an infrared divergence. However,  unlike ultraviolet divergences, infrared divergences are not a matter of renormalization but contain important physics. The spectrum of tachyons  has a specific structure dictated by the internal consistency of string theory. Experience suggests that it is wise to pay attention to what the string is trying to tell us.

To better understand the physics of the tachyons,  we note 
that the Euclidean plane  $\mathbb{R}^2$ can be regarded as the Euclidean Rindler space. 
The spacetime partition function $\hat Z_q(N)$ can be viewed as the thermal partition function for a Rindler observer:
\be
\hat Z_q(N) :=  \Tr \left[ \exp \left({-\frac{2\pi}{N} H_{R}}\right) \right]
\ee
 where  $H_{R}$ is  the Rindler Hamiltonian which generates the translations of Euclidean Rindler time corresponding to rotations in the plane. One can regard $H_R$ as  the modular  Hamiltonian of a density matrix  $\rho := \exp (-2\pi H_R)$. One is forced to define the density matrix in this indirect way since at present one does not know how to define notions such as partial trace or algebra of local observables in string theory. The partition function $\hat Z_q(N)$ can then be viewed 
as a generalization of the R\'enyi entropy:
\begin{equation}
	\hat Z_q(\cN) := \Tr (\rho^{\cN} )
\end{equation}
 for non-integer $\cN$ with $\cN = 1/N$.  On physical grounds one expects good analytic behavior of this function in the right half-plane  $\Re (\cN) \geq 1$ but not necessarily in the left half-plane $\Re (\cN) < 1$ where the tachyons exist. 
 
We elaborate on this point following the observations in  \cite{Witten:2018xfj}. In quantum mechanics, the density matrix $\rho$ is a positive Hermitean matrix normalized to $\Tr (\rho) =1$.  The eigenvalues of $\rho$ have to be less than or equal to unity. Therefore, the trace $\Tr (\rho^{\cN})$ exists in the region $\Re (\cN) \geq 1$, and the absolute value of $\hat Z_q(\cN)$ is bounded by unity. On the other hand,  $\Tr (\rho^{\cN})$ in general need not be well-defined in the region  $\Re (\cN) < 1$ --- {the convergence of $\sum_n^\infty{\lambda_n}$ does not guarantee the convergence of  $\sum_n^\infty{\lambda_n}^\cN$.} In other words, the tachyons may not be a physical threat despite their menacing comportment in the unphysical realm. 

In string theory, the partition function \eqref{Zq} is represented as an integral over a moduli space. This representation provides a useful refinement. One could ask if the tachyonic terms in the \textit{integrand}  can be summed and analytically continued to obtain  a finite and sensible physical answer for the \textit{integral} in the physical region. As we shall see,  this is indeed the case.

\subsection{Entanglement Entropy in Type-II String Theory \label{sec:Entanglment}}

Following \cite{Dabholkar:1994ai} we consider orbifolds of Type-II superstring in light-cone gauge on  flat space $\bR^6 \times \bR^2$. 
The orbifold group $\bZ_{N}= \{1, g, \ldots, g^{N-1}\}$ is generated by
\be\label{twist}
g := \exp {\frac{4\pi \mi J}{N}}
\ee
where $J$ is the generator of rotations in $\bR^2$.
The one-loop partition function can be written compactly \cite{Dabholkar:1994ai,Dabholkar:2022mxo} as an integral over the fundamental domain $\cD$ of the modular group $SL(2, \mathbb{Z})$ in the upper half $\tau$ plane:
\be\label{one-loop}
Z^{(1)}(N) = \frac{A_{H}}{N} \int_\cD  \frac{\mathrm{d}^{2}\t}{\t_{2}^{5}} 
\, \sum_{k,  \ell = 0}^{N-1}
  \,  \bigg|G \Big(\frac{k}{N} \tau + \frac{\ell}{N} \, | \t \Big)\bigg|^2 \, .
\ee
Here  $A_{H}$ is the regularized horizon area in string units 
\begin{equation}
	A_H := \frac{V_8}{(2\pi l_{s})^8} \, ,\qquad 	 \qquad l_{s}^{2}= \apm. 
\end{equation} 
The function 
\begin{equation}\label{Gdef}
	G(z|\tau) \, := \,
 \frac{1}{\eta^9(\tau)} \, \frac{\vt^4(z|\tau)}{\vt(2z|\tau)} 
 \end{equation}
 is defined in terms of the Jacobi theta-function and the  Dedekind eta-function with product representations:
 \begin{align}
	\vt (z| \t) &= -2 q^{\frac{1}{8}} \, \sin{\pi  z}
\prod_{n=1}^{\infty} (1-q^{n}) (1 - q^{n} y )\, 
 (1 - q^{n} y^{-1})\, ,  \non \\
 \eta(\tau)  &=  q^{\frac{1}{24}} 
\prod_{n=1}^{\infty} (1-q^{n})\, , \label{thetaproduct}
\end{align}
where  $q:= \exp(2\pi \mi \tau)$ and $y:= \exp (2\pi  \mi z)$. With $\tau = \tau_1 + \mi \tau_2$, the fundamental domain $\cD$ can be taken to be the usual `keyhole' region $|\tau| \geq 1$,  $\tau_2 >0$, $|\tau_1 | \leq 1/2$.

It is useful to regard the partition function as a trace over a Hilbert space of the worldsheet theory to see more clearly its physical interpretation in terms of  states. 
The orbifold Hilbert space has $N$ sectors labeled by the `twists' $k= 0, \ldots, N-1$, where $k=0$ corresponds to the untwisted sector. 
A term  with a given $k$ and $\ell$ in \eqref{one-loop} can be viewed as a trace over the oscillator modes in $k$-twisted sector of the form
\begin{equation} 
	\bigg|G\Big(\frac{k}{N} \tau + \frac{\ell}{N} \, | \t\Big)\bigg|^2 = \TrH[k]\left[g^\ell \, q^{N_L + \e_L } \, \bar q^{N_R + \e_R }\right], \label{traceg}
\end{equation}
where $N_L$ and $N_R$ are the oscillator energy operators and $\e_L$ and $\e_R$ are the ground state energies for the left and the right movers respectively in  the $k$-twisted sector. In  the Green-Schwarz light-cone formalism $\e_L = \e_R = -k/N$ in the $k$-twisted sector \cite{Dabholkar:1994ai}. The integral over $\tau_1$ in \eqref{one-loop} ensures that only level-matched states with $N_L= N_R$ contribute to the trace. 
Summing over the `twines' $\ell = 0, \ldots, N-1$ inserts the projection operator 
into the trace which ensures that only $\mathbb{Z}_N$-invariant states contribute. 

For spacetime interpretation, we recall 
that 
\begin{equation} \label{massop}
	M^2 = N_L + N_R + \e_L + \e_R \, .
\end{equation}
can be identified as the mass operator of states in the string spectrum. 
States for which $M^2$ is positive, zero, negative are respectively massive, massless, or tachyonic. One can identify $2\pi \t_2$ with the Schwinger parameter or the proper time of particle trajectories in spacetime. Large $\tau_2$ corresponds to the infrared regime whereas small $\tau_2$ corresponds to the ultraviolet regime. 

We write the  integral \eqref{one-loop} as a modular integral of a modular function $\cF(\t)$ with the Weil-Petersson measure over the fundamental domain $\cD$:
\be\label{one-loop1}
Z^{(1)}(N) = A_{H} \int_\cD  \frac{\mathrm{d}^{2}\t}{\t_{2}^{2}} \cF (\tau, N) \, .
\ee
The famously soft ultraviolet behavior of strings corresponds to the fact that the modular integral is restricted to the keyhole region. As a result, one does not expect any UV divergences which arise in field theory  from the proper time integral for the heat kernel near $\tau_2$ going to zero.
On the other hand, the integral does suffer from IR divergences because  the integrand grows exponentially as $\tau_2$ goes to infinity. 

To analyze these divergences systematically, we note that the modular function admits a Fourier expansion
\begin{equation}
	\cF(\tau, N) = \sum_n \cF_n (\tau_2, N) e^{2\pi \mi n \tau_1} \, .
\end{equation}
The zero mode is given by
\begin{equation}
	\cF_0(\t_2, N) := \frac{1}{N\t_2^3} \sum_{k=0}^{N-1} \sum_{\ell=0}^{N-1}\int_{-\half}^{\half} \mathrm{d}\t_1 \bigg|G\Big(\frac{k}{N} \tau + \frac{\ell}{N} \, | \t\Big)\bigg|^2\, .
\end{equation}
The infrared divergence comes from terms in $\cF_0(\t_2, N)$ that grow exponentially as $\t_2$ becomes large which correspond to the propagation of tachyonic states. The integral over $\t_1$ and the sum over twines $\ell$ ensure that only level-matched  and $\mathbb{Z}_N$-invariant tachyons contribute.

There are only a finite number of such terms.  The expression \eqref{traceg} is invariant  under the exchange of $k$ and $N-k$. Using this symmetry one can restrict the attention to $1 \leq k \leq \frac{N-1}{2}$. Using \eqref{Gdef} and the product representations it is easy to see that  the tachyonic part of the integrand $\cF_0^T(\t_2, N)$ can be written as
 \begin{equation}\label{Fdef}
	\cF_0^T(\t_2, N) =  \frac{2}{\t_2^3}\sum_{k=1}^{\frac{N-1}{2}} \exp \left(\frac{4\pi \tau_2k}{N} \right) f_k(\t_2, N) ,
	\end{equation}
with  $f_k(\t_2, N)$ given by
\begin{equation}
 \label{fdef}
f_k(\tau_2, N) = \sum_{r=0}^{r_k} e^{ -   r \frac{(2N - 4k ) }{N} 2 \pi \tau_2    } \,  
\end{equation}
where $r_k$ is a non-negative integer such that \eqref{Fdef} contains only tachyons. In other words, for a given $k$ in the sum \eqref{Fdef}, $r_k$ is the largest non-negative integer such  that $ r (2N -4k) < 2k$. 

It is noteworthy that all terms in the sum in \eqref{Fdef} and \eqref{fdef} have unit coefficients indicating that  tachyonic states of a given $k$ and a given mass-squared have unit degeneracy. This important fact can be easily verified also in the Hamiltonian formalism. 
 In the $k$-twisted sector, the ground state is unique with negative energy $\e_L +\e_R = -2k/N$.  In spacetime it corresponds to a tachyon  with mass-squared $M^2 = -2k/N$ and unit degeneracy. We refer to it as the `leading' tachyon  in the sense that it has the most negative $M^2$. Raising  operators  on the ground state in each sector can only increase the total energy and can give rise to subleading tachyons  as long as $M^2$ remains negative. After acting with a  sufficient number of raising operators, $M^2$ eventually becomes positive and therefore only a finite number of tachyonic terms are possible.
 It turns out that only the fractionally-moded oscillators of the single complex boson coordinatizing the Rindler plane $\bR^2$ are relevant because the action of raising operators of other fields yield nontachyonic states. As a result, the subleading tachyons also have unit degeneracy.

\subsection{Analytic Continuation of Tachyonic Contributions \label{sec:Analytic}}

The spectrum is thus replete with tachyons and the  partition function \eqref{one-loop} is badly divergent in each twisted sector for all $k=1, \ldots, N-1$. It seems hopeless to try to make sense of the integral for $N > 1$. Remarkably, the precise structure of string theory allows for a summation of the tachyonic  terms in the integrand. This sum  can be analytically continued to the physical region $\Re(\cN) \geq 1$ where it tends to a finite limit as $\t_2$ tends to infinity and the modular integral \eqref{one-loop2} is then finite. 

The function $f_k(\tau_2, N)$ depends on $r_k$ with an intricate dependence on $k$ which makes it difficult to obtain a simple answer for the sum over $k$ in \eqref{Fdef}. It is useful to consider instead a function $\tilde f_k (\tau_2, N)$ obtained by taking $r_k$ to infinity in \eqref{fdef}. It corresponds to adding  infinitely many (fictitious) massive  string states and some massless ones which do not change the convergence properties at large $\tau_2$.  One can similarly define $\tilde \cF^T_0(\tau_2, N)$ by replacing $f_k (\tau_2, N)$ by $\tilde f_k (\tau_2, N)$ in \eqref{Fdef}. The $k$-sum  can now be readily performed for all values of $\tau_2$ to obtain
\begin{equation}
\label{sumFT2}
	\tilde \cF^T_0(\t_2, N) = -\frac{2}{\t_2^3} \sum_{r=0}^\infty e^{- 4r \pi \tau_2}\frac{1-e^{\frac{\pi  (N-1) (4 r+2) \tau _2}{N}}}{1-e^{-\frac{2 \pi  (4 r+2) \tau _2}{N}}} \, .
\end{equation}
Remarkably, this function  is perfectly finite for $0< N \leq 1$ as $\tau_2\to \infty$ even though it diverges for $N >1$.

To separate the tachyonic and non-tachyonic contributions, one can rewrite the partition function \eqref{one-loop1} as 
\be\label{one-loop2}
Z^{(1)}(N) = A_{H} \int_\cD  \frac{\mathrm{d}^{2}\t}{\t_{2}^{2}} \left[ \tilde \cF^R (\tau, N) + \tilde \cF^T_0 (\t_2, N) \right]\, 
\ee
where $\tilde \cF^R (\tau, N)$ is the nontachyonic remainder of $\cF (\tau, N)$ defined by  $\tilde \cF^R (\tau, N):= \cF (\tau, N) - \tilde \cF^T_0 (\t_2, N) $. 
With this splitting, one can examine the dependence on $N$ of  each of the two terms in \eqref{one-loop2} separately. 

The integral of $\tilde \cF^R(\t, N)$ has no tachyonic divergences by construction and is convergent both in the IR and the UV. This finite integral could be performed numerically for each $N$ to obtain finite values for all odd integers. An interpolation method like the Newton Series \cite{Dabholkar:2022mxo} could then be used to deduce from this data an analytic continuation or `extrapolation' in the physical region $0 < N \leq 1$ as long as there is no unexpected oscillatory behavior. 

The integral of $ \tilde \cF^T_0 (\t_2, N)$ is divergent in the region $N >1$ but  has a finite limit as $\tau_2$ tends to infinity for $ 0 < N \leq 1 $. We thus find that the tachyonic part of the integrand can be summed and analytically continued such that the total integral is free of the IR divergences in the physical domain $0 < N \leq 1$ or $\cN \geq 1$. 

It is worth emphasizing that this surprising finiteness  is not accidental but depends critically on three very specific `just so' properties of superstring theory. 
\begin{enumerate}
	\item There are exactly $N-1$ leading tachyons with unit degeneracy in each twisted sector.  The analytic continuation of \eqref{sumFT2}  would not have the desired behavior in the $\cN$ plane if there were, for example, $N+1$ leading tachyons, or if the multiplicities were different. 
	\item The total ground state energy $-2k/N$ in the $k$-twisted sector is linear in $k$.  In the light-cone Green-Schwarz formalism, there are four complex fermions twisted by $k/N$ with ground state energy  $+1/12   - (1-k/N)k/2N $. There is   single complex boson twisted by $2k/N$ with ground state energy $-1/12 + (1-2k/N)k/N $ for $2k < N$, and three untwisted complex bosons with ground state energy $-1/12$. The contributions quadratic in $k/N$ cancel out as do the constant terms. This precise cancelation resulting in ground state energy linear in $k$ depends on the specific structure of the superstring and is not true, for example, for the bosonic string. 
	\item The subleading tachyons also have energies linear in $k$ and have unit degeneracies for reasons explained earlier. A simple geometric sum \eqref{sumFT2}  would not be possible without these properties. 
\end{enumerate}
   
The taming of the tachyons in the physical region is very encouraging. One can now compute the entropy from the region near $N =1$ approaching from $N <  1$. It is interesting that one can extract a nontrivial finite answer for a physical quantity even though in the intermediate steps supersymmetry is broken and the spectrum is inflicted with  tachyons. A similar phenomenon has been noted in the open string sector using quite different methods \cite{Witten:2018xfj} and appears to be a general feature.  

There is a perverse possibility that the terms coming from massless and massive modes which are finite for $N>1$, sum up to a divergent answer for $0 < N \leq 1$. On physical grounds this seems unlikely. For any set of massive or massless fields, the physical density matrix $\rho$ is expected to have eigenvalues  less than or equal to unity. With the ultraviolet cutoff provided by string theory, it would be unphysical if $\Tr (\rho^\cN )$ turns out to be convergent for  $\cN < 1$ but divergent for $\cN \geq 1$. In any case, it is important to explore the integral further to rule out this eventuality and explicitly compute the finite remainder. 

The situation is reminiscent of the Euler gamma function $\Gamma(N)$. The representation of a function  analytic in the entire $N$ plane which reduces to the factorial for positive integers can be deduced knowing the values of the function for  positive integers and its integral representation in the region $\Re(N) > 0$. The resulting analytic function has poles at non-positive integers in the region $\Re(N) \leq 0$. The situation for the function $Z(N)$ is somewhat similar but, in fact, in the reverse.  We seem to be required to determine the analytic representation of $Z(N)$ knowing something about the function in the region $\Re (N ) > 1$ for odd integers where the function is ill-defined because of the tachyonic divergence. 

Fortunately, string theory provides an integral representation for this function. 
Even though the integral \eqref{one-loop2} diverges for $N > 1$, the integrand is finite for fixed $\tau$. The tachyonic part is easy to analytically continue to the physical region.  Our results indicate that the resulting integral in the region $N \leq\ 1$ is finite, which is an essential ingredient in obtaining a sensible physical interpretation.

\subsection{Discussion}

The orbifold method thus seems to yield a finite answer for the entropy defined by \eqref{S-total} to one-loop order. There is no renormalization of Newton's constant in the ten-dimensional superstring. Therefore, in this case the finite part of our one-loop computation gives directly the one-loop entanglement entropy. 
 One expects to be able to garner more information than just this single number. The analyticity property suggests that one can obtain a finite expression for the entropy $\Tr (\rho^\cN)$ as an analytic function of $\cN$ in the physically relevant region $\Re (\cN) \geq 1$ to learn more about the short-distance degrees of freedom of string theory. The fact that $Z(1)=0$ is consistent with this interpretation. Using \eqref{spaceworld} it suggests that $\Tr (\r) =1$, and therefore it may be possible define a density matrix with a well-defined trace.  Finiteness of the entanglement entropy is physically very significant for various reasons which are worth recalling. 

As discussed earlier, the resolution of the information paradox is closely linked to the finiteness of the entanglement entropy. Unitarity of the boundary theory in holography indicates  that most probably the time evolution in the bulk is  also unitary. However, it is essential to understand the resolution of the information paradox directly in the bulk gravitational theory. Understanding the finiteness of entanglement is a  step towards this goal. Entanglement entropy is  a critical ingredient also in the formulation of the strong sub-additivity paradox for Hawking emission \cite{Mathur:2005zp, Mathur:2009hf,  Almheiri:2012rt},
 or in the proof of the generalized second law of thermodynamics \cite{Wall:2011hj}. Holographic entanglement entropy beyond the classical formula also requires a definition of entanglement in the bulk \cite{Ryu:2006bv,  Hubeny:2007xt,  Lewkowycz:2013nqa,  Barrella:2013wja,  Faulkner:2013ana,  Engelhardt:2014gca,  Jafferis:2015del}; and quantum entanglement entropy is relevant for defining the quantum extremal surface and the proposals for reconstructing bulk observables and the black hole interior \cite{Hamilton:2005ju, Hamilton:2006az, Papadodimas:2012aq,  Papadodimas:2013wnh, Almheiri:2014lwa,  Harlow:2018fse,  Jafferis:2020ora}. {Some of the works implementing the replica method assume absence of tachyons --- it would be interesting to explore this in more detail.}
A string-theoretic definition of finite entanglement entropy is  desirable in these various contexts.  
 
As emphasized in \cite{Dabholkar:2022mxo}, entanglement entropy in flat space computed here is relevant  near the horizon of any large mass two-sided black hole with bifurcate horizon, in particular, the Schwarzschild black hole in anti de Sitter spacetime. In this context, the ground state of the bulk spacetime theory corresponds in the boundary to an entangled state in the thermofield double \cite{Maldacena:2001kr, VanRaamsdonk:2010pw, Maldacena:2013xja}. Tracing over the left conformal field theory results in a thermal state in the right conformal field theory. The entanglement entropy of the thermo-field double state  thus equals the entropy of the thermal bath, which has a systematic double expansion that can be matched to a perturbative expansion in the bulk in the string coupling $g_s$ and the string scale $l_s^2$. The classical Bekenstein-Hawking entropy corresponds to the leading entropy of the thermal state of order $N^2$ in the large $N$ limit.  One expects that the quantum entanglement entropy can be matched to the order $N^0$ correction  which is finite and in principle calculable. It would be interesting to make this comparison for the finite mass, nonsupersymmetric black hole.

In the holographic context, the algebra of observables of the right CFT is manifestly Type-I since it admits an irreducible representation over the right Hilbert space, reflected in the fact that the entanglement entropy is finite. One expects a corresponding statement in the bulk as discussed in \cite{Dabholkar:2022mxo}. The algebra of observables is Type-III in the field theory limit of the bulk but one expects that string theory should ameliorate the situation. Finiteness of entanglement indicates that the algebra of observables of the bulk quantum gravity is indeed akin to Type-I as in the boundary theory.  In quantum gravity one cannot really define an algebra of local observables and it is not clear what generalization will correspond to notions like modular Hamiltonian and  entanglement entropy. See \cite{Leutheusser:2021qhd,Leutheusser:2021frk,Witten:2021unn,Chandrasekaran:2022eqq,Chandrasekaran:2022cip,Witten:2023qsv} for recent discussions.  
A computable and  finite entanglement entropy in string theory would be a useful guide in  the search towards such a generalization. The orbifold has an exact conformal field theory description  which is the required data for defining off-shell string field theory on the conical background. It would be interesting if the machinery of string field theory \cite{deLacroix:2017lif,  Erbin:2021smf} can be brought to bear on this important problem.

\vskip 3mm
\centerline{ \textit{Acknowledgements}}
\vskip 3mm
{
\noindent
We thank Don Zagier for useful conversations.
}

\bibliographystyle{JHEP}
\bibliography{entangle}

\end{document}